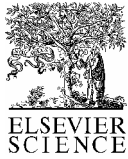
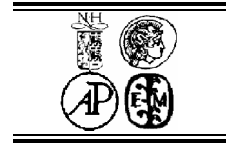

# Experimental realization of Laughlin quasiparticle interferometers

F. E. Camino, Wei Zhou, V. J. Goldman[*]

*Department of Physics, Stony Brook University, Stony Brook, NY 11794-3800, USA*

**Abstract**

Laughlin quasiparticles are the elementary excitations of a highly-correlated fractional quantum Hall electron fluid. They have fractional charge and obey fractional statistics. The quasiparticles can propagate quantum-coherently in chiral edge channels, and constructively or destructively interfere. Unlike electrons, the interference condition for Laughlin quasiparticles has a non-vanishing statistical contribution that can be observed experimentally. Two kinds of interferometer devices have been realized. In the primary-filling interferometer, the entire device has filling 1/3, and the $e/3$ edge channel quasiparticles encircle identical $e/3$ island quasiparticles. Here the flux period is $h/e$, same as for electrons, but the back-gate charge period is $e/3$. In the second kind of interferometer, a lower density edge channel at filling 1/3 forms around a higher density island at filling 2/5, so that $e/3$ edge quasiparticles encircle $e/5$ island quasiparticles. Here we observe superperiodic oscillations with $5h/e$ flux and $2e$ charge periods, both corresponding to excitation of ten island quasiparticles. These periods can be understood as imposed by the anyonic braiding statistics of Laughlin quasiparticles. © 2001 Elsevier Science. All rights reserved

A clean system of 2D electrons subjected to high magnetic field at low temperatures condenses into the fractional quantum Hall (FQH) fluids. [1-3] An exact filling $f$ FQH condensate is incompressible and gapped. The celebrated examples of FQH condensates are the Laughlin many-electron wave functions for the *primary* fillings $f = 1/(2j + 1)$, with $j$ a positive integer. The elementary charged excitations of a FQH condensate are the Laughlin quasiparticles. Deviation of the filling factor from the exact value is achieved by excitation of either quasielectrons or quasiholes out of the condensate. At such fillings, the ground state of a FQH fluid consists of the quasiparticle-containing condensate. The FQH quasiparticles have fractional electric charge [2-4] and obey anyonic exchange statistics. [5-9]

In 2D, the laws of physics permit existence of particles known as anyons, whose statistics $\Theta$ is intermediate to Bose and Fermi. [10,11] When two particles of a system of bosons are exchanged, the phase of the system remains unchanged, and for a system of fermions it changes by $\pi$. Swapping two anyons results in a phase factor intermediate between zero and $\pi$, the precise value being characteristic of the species of the particles. Just as the quantum statistics of bosons and fermions gives rise to Bose-Einstein condensation and the Pauli exclusion principle, the more exotic behavior of anyons has profound implications.

Fractionally charged FQH quasiparticles were first observed directly in quantum antidot experiments. [4,12,13] A quantum antidot is a small potential hill, defined lithographically in the 2D electron system. Quasiperiodic resonant conductance peaks are observed when the quasiparticle occupation of the antidot is incremented one at a time. Complementary device geometry, where an electron island is separated from the 2D system by two nearly open constrictions, is an interferometer. [7-9,14-19]

In this paper we compare results obtained from two interferometer devices. Small differences in device design and fabrication have led to one having more depleted constrictions than another. This resulted in a dramatic difference in device performance in the FQH regime. Not long ago, we reported [7,8,18] realization of a quasiparticle interferometer where $e/3$ quasiparticles of the 1/3 fluid execute a closed path around an island of the 2/5 fluid. The interference fringes were observed as conductance oscillations as a function of the magnetic flux through the island, like the Aharonov-Bohm effect. The observed flux and charge periods, $5h/e$ and $2e$, are equivalent to excitation of ten $e/5$ quasiparticles of the 2/5 fluid. The superperiod is interpreted as imposed by the anyonic statistical interaction of the quasiparticles. [6,20,21]

Recently, we reported [9] a comparable quasiparticle interferometer, but with much less depleted constrictions, Fig. 1. This results in the entire electron island being at the primary quantum Hall filling 1/3 under coherent tunneling conditions. In this regime, the interfering $e/3$ quasiparticles execute a closed path around an island of the 1/3 FQH fluid containing other $e/3$ quasiparticles. The observed flux and charge periods, $h/e$ and $e/3$ respectively, correspond to addition of one quasiparticle to the area enclosed by the interference path. These periods are the same as in quantum antidots, but the quasiparticle path encloses no electron vacuum in the interferometer. This is consistent with the Berry phase quantization condition that includes both Aharonov-Bohm and anyonic statistical contributions.

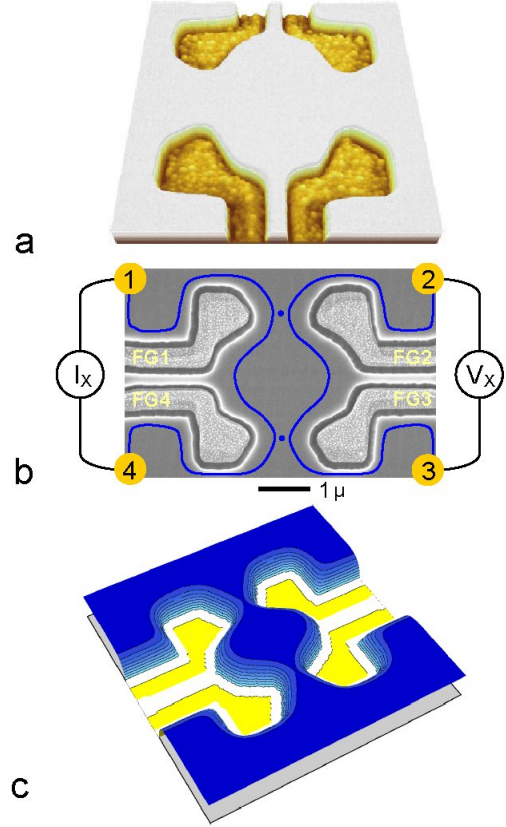

FIG. 1. The primary filling quasiparticle interferometer device. Atomic force (a) and scanning electron (b) micrographs. Four front gates (FG) are deposited in shallow etch trenches. Depletion potential of the trenches defines an electron island with lithographic radius 1.34 μm. The chiral edge channels (blue lines) follow an equipotential at the periphery of the undepleted 2D electrons. Tunneling (blue dots) occurs at the saddle points in the two constrictions. The edge channel path encircling the island is closed by tunneling, forming the interferometer. The back gate (not shown) extends over the entire 4×4 mm sample. (c) Illustration of the 2D electron density profile of the primary-filling interferometer device. Conductance oscillations are observed when the counterpropagating edge channels are coupled by tunneling at the saddle points in the constrictions. The edge ring density is 0.93 of the island center density, and the entire interferometer has the same quantum Hall filling for wide plateaus, such as $f = 1$ or 1/3.

The interferometer devices were fabricated from low disorder AlGaAs/GaAs heterojunctions. The four independently-contacted front gates were defined by electron beam lithography on a pre-etched mesa with Ohmic contacts. After a shallow 130 - 180 nm wet etching, Au/Ti front-gate metal was deposited in the etch trenches, followed by lift-off, Fig. 1(a,b). Even at zero front-gate voltage, the GaAs surface depletion of the etch trenches creates electron confining potential defining two wide constrictions, which separate an approximately circular electron island from the 2D "bulk".

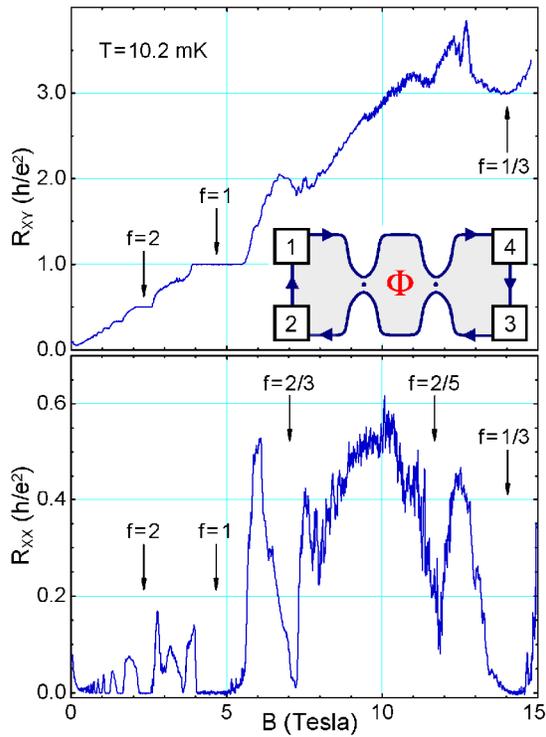

FIG. 2. The diagonal $R_{XX}$ and Hall $R_{XY}$ resistance of the interferometer device at zero front-gate voltage for the device of Fig. 1. The quantized plateaus allow to determine the filling factor in the constrictions. The fine structure is due to quantum interference effects in the residual disorder potential and the interferometric conductance oscillations as a function of magnetic flux through the island. Inset: the chiral edge channel electron interferometer concept; dots show tunneling.

Samples were mounted on sapphire substrates with In metal, which serves as the global back gate, and were cooled to 10.2 mK in a dilution refrigerator. Extensive cold filtering cuts the electromagnetic environment incident on the sample, allowing to achieve electron temperature ≤15 mK in an interferometer device. [19] Four-terminal resistance $R_{XX} = V_X/I_X$ was measured with 50 pA ($f = 1/3$ regime) or 200 pA ($f = 1$ regime), 5.4 Hz AC current. The Hall resistance $R_{XY}$ is determined by the quantum Hall filling $f$ in the constrictions, giving definitive values of $f$. The oscillatory conductance $\delta G$ is calculated from the directly measured $\delta R_{XX}$ and the quantized Hall resistance as $\delta G = \delta R_{XX}/R^2_{XY}$, a good approximation for $\delta R_{XX} << R_{XY}$.

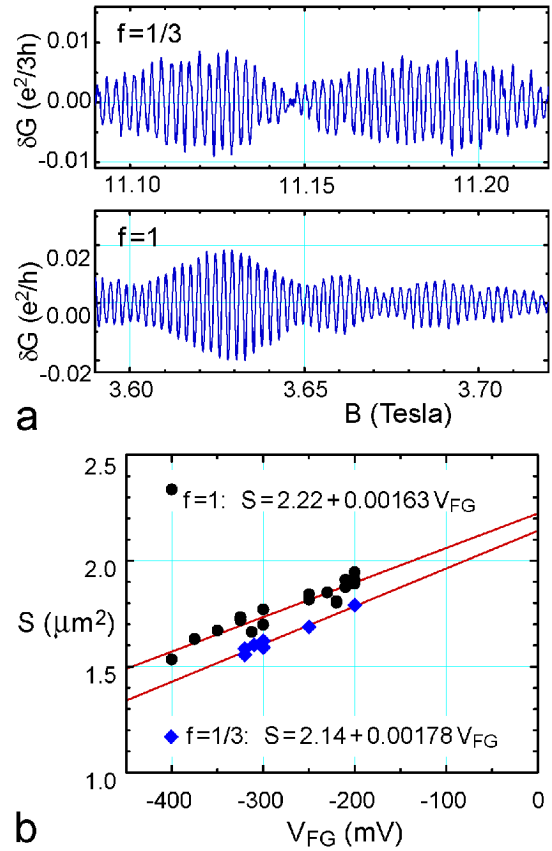

FIG. 3. (a) Representative interference conductance oscillations for electrons ($f = 1$) and for $e/3$ quasiparticles ($f = 1/3$). Moderate $V_{FG} \approx -250$ mV is applied to increase the oscillation amplitude; this reduces the island electron density and shifts the region of oscillations to lower $B$. (b) Interference path area $S = h/e\Delta_B$ as a function of front-gate voltage. Extrapolated to $V_{FG} = 0$, the magnetic field periods $\Delta_B$ are equal within the experimental uncertainty. Thus the flux period is $h/e$ in both regimes.

The etch trenches define two ≈1.2 μm wide constrictions, which separate an approximately circular electron island from the 2D bulk. Moderate front-gate voltages $V_{FG}$ are used to fine tune the constrictions for symmetry of the tunneling amplitude and to increase the oscillatory interference signal. The $B = 0$ shape of the electron density profile is primarily determined by the etch trench depletion, illustrated in Fig. 1(c). For the 2D bulk density $n_B = 1.25 \times 10^{11}$ cm$^{-2}$, there are ~3,500 electrons in the island. The depletion potential has saddle points in the constrictions, and so has the resulting density profile.

The counterpropagating edge channels pass near the saddle points, where tunneling occurs. In the range of $B$ where the interference oscillations are observed, the filling of the edge channels is determined by the saddle point filling. [18] This allows to determine the saddle point density from the magnetotransport, Fig. 2. The Landau level filling $\nu = hn/eB$ is proportional to the electron density $n$, accordingly the constriction $\nu$ is lower than the bulk $\nu_B$ in a given $B$. In the device of Fig. 1, the island center $n$ is estimated to be 3% less than $n_B$, the constriction - island center density difference is ~7%. Thus, the whole island is on the same plateau for strong quantum Hall states with wide plateaus, such as $f = 1$ and 1/3. While $\nu$ is a variable, the quantum Hall exact filling $f$ is a quantum number defined by the *quantized* Hall resistance as $f = h/e^2 R_{XY}$.

In the integer quantum Hall regime, the Aharonov-Bohm ring is formed by the two counter-propagating chiral edge channels passing through the constrictions. Electron backscattering, completing the interference path, occurs by quantum tunneling at the saddle points in the constrictions, Fig. 1. The relevant particles are electrons of charge $-e$ and Fermi statistics, thus we can obtain an absolute calibration of the interferometer path area and the back-gate action of the interferometer. Figure 3 shows the oscillatory conductance $\delta G$ calculated from the $R_{XX}$ data after subtracting a smooth background. The smooth background has two contributions: at $\nu_B$ outside the bulk plateau regions, the bulk conduction, and the non-oscillatory part of the tunneling conductance in the interferometer. Extrapolated to $V_{FG} = 0$, [16,18] the $f = 1$ magnetic field oscillation period is $\Delta_B = 1.86$ mT. This gives the interferometer path area $S = h/e\Delta_B = 2.22$ μm$^2$.

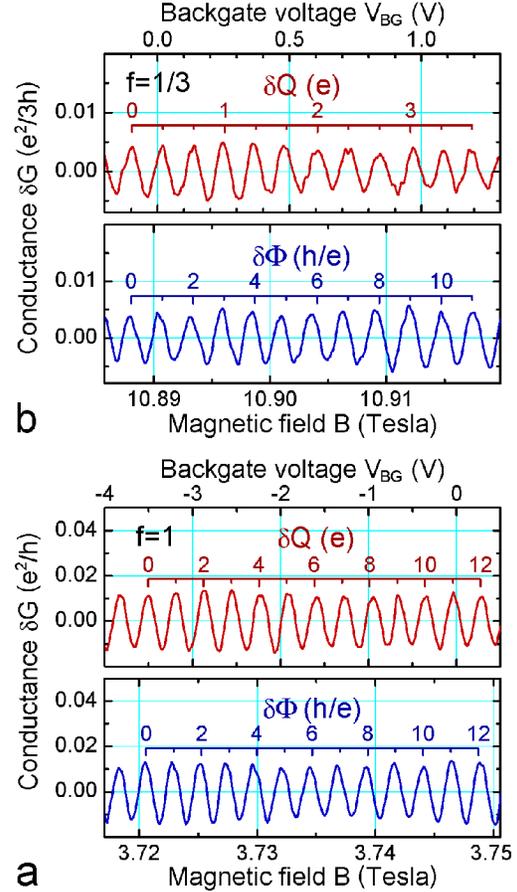

FIG. 4. Matched sets of oscillatory conductance data giving the $e/3$ charge period. (a) The interferometer device is calibrated using the conductance oscillations for electrons, $f = 1$. (b) This calibration gives the charge for the Laughlin quasielectrons $q = 0.33e$. The magnetic flux period $\Delta_\Phi = h/e$, the same in both regimes, implies anyonic statistics of the fractionally charged quasiparticles.

In the $f = 1/3$ FQH regime, we observe the interferometric oscillations as a function of magnetic field, Fig. 3. Extrapolated to $V_{FG} = 0$, the magnetic field oscillation period is $\Delta_B = 1.93$ mT. Assuming the flux period is $h/e$, this gives the interferometer path area $S = h/e\Delta_B = 2.14$ μm$^2$. When oscillations are seen, the island edge ring area is strictly determined by the condition that the edge channels pass near the saddle points in the constrictions. Classically, increasing $B$ by a factor of ~3 does not affect the electron density distribution in the island at all. Quantum corrections are expected to be small for a large island with ~3,500 electrons. Indeed, the $f = 1/3$ interferometer path area is within ±3% of the integer value. Because in the integer regime the oscillations have an $h/e$ fundamental flux period, we conclude that the flux period of the 1/3 FQH oscillations must be $\Delta_\Phi = h/e$, too.

We use the back-gate technique [4,12] to measure directly the charge period in the fractional regime. We calibrate the backgate action $\delta Q/\delta V_{BG}$, where $Q$ is the electronic charge within the interference path. The calibration is done by evaluation of the coefficient α in

$$\Delta_Q = \alpha[\Delta(V_{BG})/\Delta_B] , \qquad (1)$$

setting $\Delta_Q = e$ in the integer regime. Note that Eq. (1) normalizes the back-gate voltage periods by the experimental $B$-periods, approximately canceling the variation in device area due to a front-gate bias. The coefficient α in Eq. (1) is known *a priori* in quantum antidots to a good accuracy because the antidot is completely surrounded by the quantum Hall fluid. [4,13] But, in an interferometer a calibration is required because the island is separated from the 2D electron plane by the front-gate etch trenches, so that its electron density is not expected to increase by precisely the same amount as $n_B$.

Figure 4 shows the oscillations as a function of $V_{BG}$ for $f = 1$ and 1/3 and the corresponding oscillations as a function of $B$. At each filling, the front-gate voltage is the same for the (vs. $V_{BG}$, vs. $B$) set. The $f = 1$ back-gate period corresponds to an increment $\Delta_N = 1$ in the number of electrons within the interference path. We obtain $\Delta(V_{BG}) = 315$ mV, $\Delta_B = 2.34$ mT, and the ratio $\Delta(V_{BG})/\Delta_B = 134$ V/T. This period ratio is 0.92 of that obtained in quantum antidots, [12] consistent with expectation. For the 1/3 FQH oscillations we obtain $\Delta(V_{BG}) = 117$ mV, $\Delta_B = 2.66$ mT, and the ratio $\Delta(V_{BG})/\Delta_B = 44.1$ V/T. Using the integer calibration in the same device, the $e/3$ quasiparticle experimental charge period is $\Delta_Q = 0.328e$, some 1.7% less than $e/3$. To the first order, using the $\Delta(V_{BG})/\Delta_B$ ratio technique cancels the dependence of the $V_{BG}$ and $B$ periods on the interferometer area and front-gate bias. The scatter of the quasiparticle charge values obtained from several matched data sets in this experimental run is ~3%.

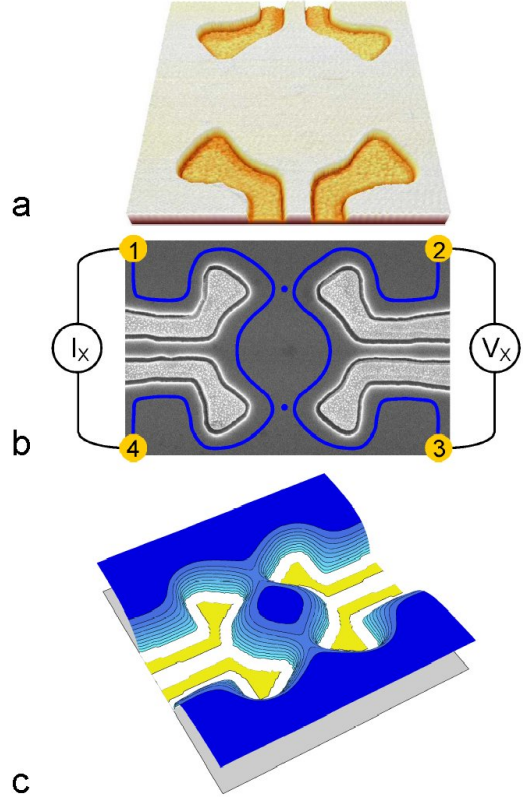

FIG. 5. The 2/5 island in 1/3 FQH Laughlin quasiparticle interferometer device. Atomic force (a) and scanning electron (b) micrographs. Four front gates are deposited in shallow etch trenches, defining an island with lithographic radius 1.05 μm. (c) Illustration of the 2D electron density profile of this interferometer device. Due to small differences in fabrication, the constrictions are ~3 times more depleted than in the device of Fig. 1. Correspondingly, the edge ring density is 0.78 of the island center density, thus allowing formation of a filling 1/3 edge ring passing through the constrictions, when the island and the 2D bulk have filling 2/5.

Another interferometer device [7,8] is shown in Fig. 5. The principal differences from the device of Fig. 1 are: (i) the smaller lithographic radius 1.05 μm; (ii) slightly narrower (by about 0.1 μm) constriction gaps, and (iii) ~20 nm deeper etch of the front-gate trenches. These differences combine to yield about three times more depleted constrictions, with the saddle point electron density estimated as 0.75 of the 2D bulk density and 0.78 of the island center density. This results in formation of a filling 1/3 edge ring passing through the constrictions, when the island and the 2D bulk both have FQH filling 2/5.

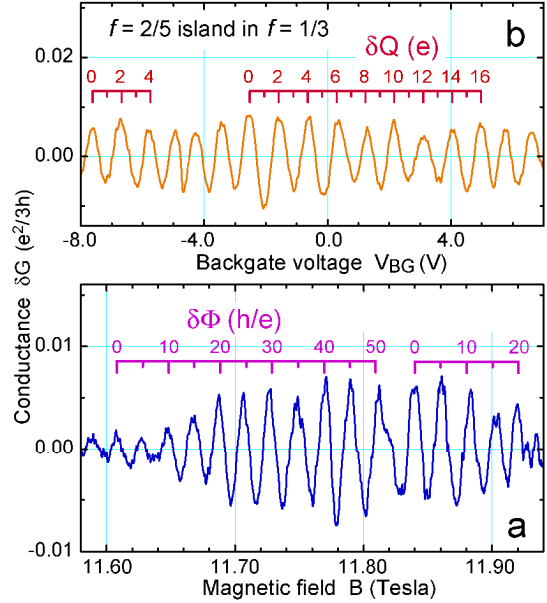

FIG. 6. Oscillatory conductance for $e/3$ quasiparticles circling an island of $f = 2/5$ FQH fluid. (a) Flux through the island period $\Delta_\Phi = 5h/e$ corresponds to creation of ten $e/5$ quasiparticles in the island [one $h/e$ excites two $e/5$ quasiparticles from the 2/5 FQH condensate, the total (quasiparticles + condensate) charge is fixed]. Such superperiod $\Delta_\Phi > h/e$ has never been reported before. (b) The charge period $\Delta_Q = 2e$ confirms that the $e/3$ quasiparticle consecutive orbits around the 2/5 island are quantized by a condition requiring increment of ten $e/5$ quasiparticles.

Figure 6 shows conductance oscillations for the device of Fig. 5. The flux and charge calibration was done in the integer QH regime, as described above. [7,8,18] The observed flux and charge periods are $\Delta_\Phi = 5h/e$, and $\Delta_Q = 2e$, respectively. Such superperiodic Aharonov-Bohm period $>h/e$ had never been reported before in any system. If we neglect the symmetry properties of the FQH condensates, the island charge could change in increments of one quasiparticle charge, any small (less than $e$) charge imbalance supplied from the contacts. Thus the superperiods must be imposed by the symmetry properties of the two FQH fluids. [20] The current used to measure conductance is transported by the quasiparticles of the outside 1/3 fluid. Therefore, the conductance oscillations periodicity of ten $e/5$ quasiparticles results from the $2\pi$ Berry phase periodicity of the charge $e/3$ Laughlin quasiparticle encircling ten quasiparticles of the $f = 2/5$ FQH fluid. [6,21]

In conclusion, we have experimentally realized two kinds of Laughlin quasiparticle interferometers, where $e/3$ quasiparticles execute a closed path around an island containing FQH fluid. The central results obtained, the flux and charge periods are robust. Both the Aharonov-Bohm flux and the charging periods are interpreted as imposed by the anyonic braiding statistics $\Theta_{1/3} = 2/3$ of the $e/3$ FQH quasiparticles.

This work was supported in part by the NSF and by U.S. ARO.